\documentclass[%
 reprint,
superscriptaddress,
 amsmath,amssymb,
 aps,
]{revtex4-1}

\usepackage[english]{babel}
\usepackage{graphicx}
\usepackage{dcolumn}
\usepackage{bm}

\usepackage{comment}
\usepackage[fleqn]{nccmath}
\usepackage{color}

\newcommand{\new}[1]{{\color{black} {#1}}}

\begin{document}



\title{Dynamical Methods for Target Control of Biological Networks}

\author{Thomas Parmer}
\affiliation{Center for Complex Networks and Systems Research, Luddy School of Informatics, Computing, and Engineering, Indiana University, Bloomington, Indiana 47408, USA}
\author{Filippo Radicchi}
\affiliation{Center for Complex Networks and Systems Research, Luddy School of Informatics, Computing, and Engineering, Indiana University, Bloomington, Indiana 47408, USA}


\begin{abstract}
Estimating the influence that individual nodes have 
on one another 
in a Boolean network is essential 
to predict and 
control
the system's dynamical behavior, 
for example, detecting key therapeutic targets to control pathways in models of biological signaling and regulation.
Exact estimation 
is generally not possible due to the fact that the number of configurations that must be considered grows exponentially with the system size. However, approximate, scalable methods exist in the literature. These methods can be divided in two main classes: (i) 
graph-theoretic methods that rely  on representations of Boolean dynamics into static graphs, (ii) and 
mean-field approaches that describe average trajectories of the system but neglect dynamical correlations. 
Here, we compare systematically the performance of these state-of-the-art methods on a large collection of real-world gene regulatory networks. We find comparable performance across 
methods. All 
methods underestimate the ground truth, with mean-field approaches having a better recall but a worse precision than 
graph-theoretic
methods. Computationally speaking, graph-theoretic methods are faster than mean-field ones in 
sparse 
networks, but are slower in dense
networks. The preference of which method to use, therefore, depends on a network's connectivity and the relative importance of recall {\it vs.} precision for the specific application at hand.
\end{abstract}

\maketitle

\section*{Introduction}

Understanding the influence that individual elements have on other elements in a complex dynamical system is essential for the prediction and control of the system's behavior.
Such  a notion of influence is studied broadly on Boolean networks.
Examples include studies concerning
perturbations~\cite{kauffman1993origins,kauffman2004proposal,serra2004genetic,ramo2006perturbation}, causal inference~\cite{klamt2006methodology,saez2007logical,samaga2010computing,wang2011elementary,marques2013canalization,zanudo2015cell}, and control~\cite{ghanbarnejad2012impact,fiedler2013dynamics,mochizuki2013dynamics,zanudo2017structure}.

Some papers focus on
the effects that pinning specific variables to invariant values has
on the rest of the system's dynamics without having knowledge of the exact configuration that the system is in~\cite{samaga2010computing,zanudo2017structure,yang2018target,parmer2022influence,parmer2023dynamical, 9528951, 9505627}. 
Pinning a set of seed nodes 
may drive other nodes to deterministic long-term dynamical states.
This set of controlled nodes is named as the domain of influence of the seed set~\cite{yang2018target}.
The identification of domains of influence is useful in target control, whereby partial knowledge of a system's state can be used to infer the state of an uncontrolled, target set of variables, e.g., driving a cancer cell towards an apoptotic state. Unfortunately, the exact determination of the domain of influence of a seed set is a computationally infeasible task because of the exponentially large number of configurations that a Boolean network can assume.

\begin{figure*}[!htb]
\centering
\includegraphics[width=1.9\columnwidth]{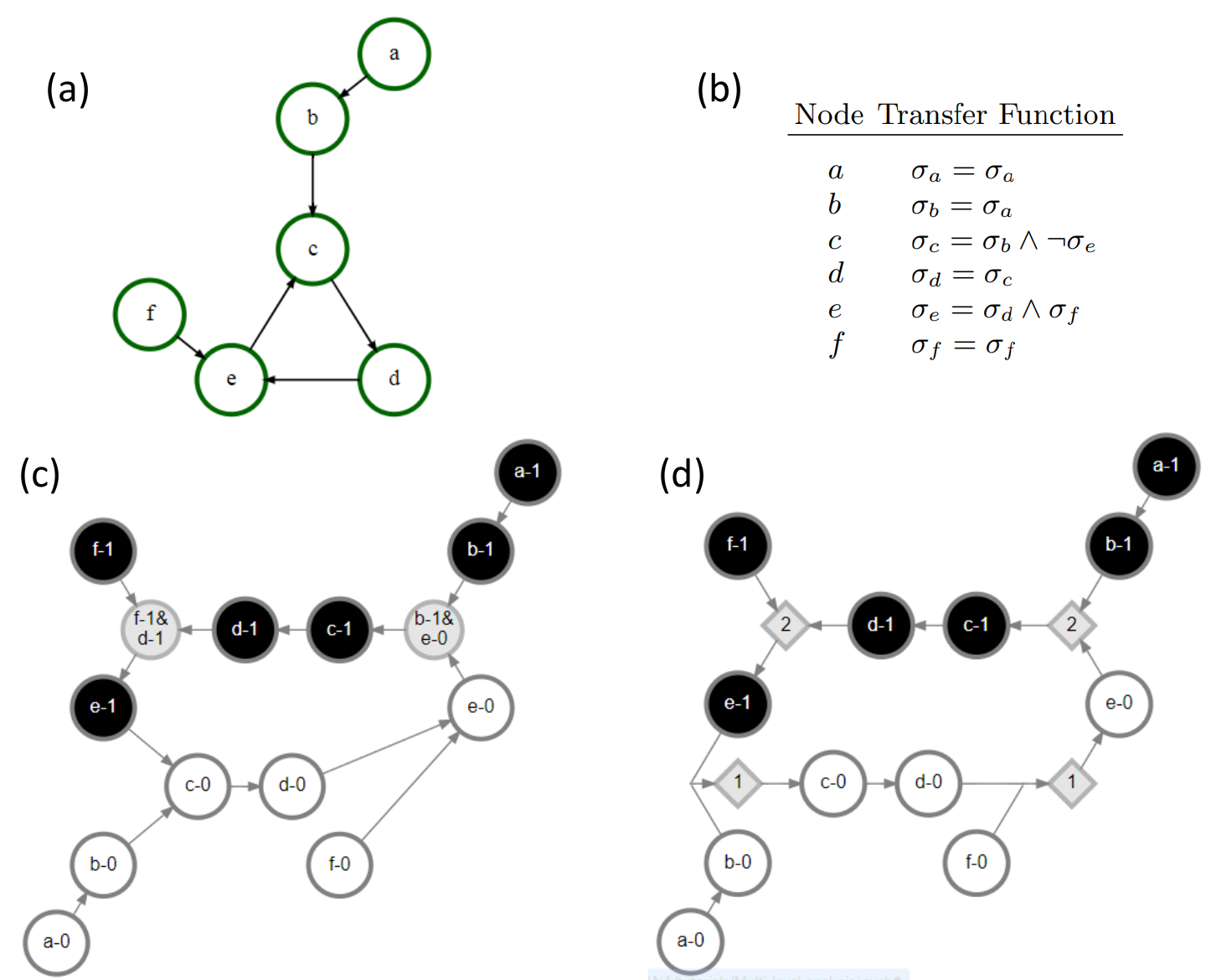} 
\caption{\textbf{Static representations of the dynamics on a Boolean network.}
(a) We consider a toy example of a Boolean network. A direct connection indicates that a variable's state depends on the state of the other variable. In this network for example, the state of node \textit{e} depends on the states of nodes \textit{d} and \textit{f}. Nodes \textit{a} and \textit{f} are instead inputs in this network, in the sense that their state is time invariant.  (b) The transfer functions are represented in logical form for each node based on the states of its neighbors. 
(c) Expanded network representation of the Boolean network. Each node of the original network of panel (a) is denoted by its label plus its state. For example, ``$e$-$1$'' indicates the state $\sigma_e=1$ of node $e$; the composite node with label ``$f$-$1 \, \& \,  d$-$1$'' denotes instead the simultaneous appearance of the states $\sigma_f=1$ and $\sigma_d=1$. In the visualization, nodes representing active states are denoted in black, nodes representing inactive states are displayed in white, and composite nodes representing AND relationships  are denoted in grey.
(d) Dynamics canalization map representation; s-units representing active nodes are denoted with black circles and s-units representing inactive nodes are denoted with white circles. T-units representing redescribed schemata are displayed with grey diamonds. The label appearing in each of the t-units represents the specific value of the threshold that they represent. T-units with threshold equal to one that represent schemata with no permutation redundancy are left out of the figure for simplicity.
Additionally, self-loops are left out of panels (a), (c) and (d).
}
\label{fig:GRN_representations}
\end{figure*}

Some approximate methods exist in the literature.
These include 
graph-theoretic 
models such as
the logical interaction hypergraph (LIH) \cite{klamt2006methodology}, the expanded network \cite{wang2011elementary}, and the dynamics canalization map (DCM) \cite{marques2013canalization}.
The three methods above provide static graphs that represent the dynamics of a Boolean network. 
Specifically, the 
LIH represents a Boolean network as a signed directed hypergraph by adding signs to each interaction and hyperarcs to represent logical AND relationships.  The 
expanded network
uses composite nodes to represent AND relationships and complementary nodes to represent variable negation for interactions involving NOT operations (Fig. \ref{fig:GRN_representations}c).  Finally, the 
\new{DCM
is a threshold network built from the minimized version of Boolean functions using Blake's canonical form \cite{blake1937canonical}, which
uses s-unit nodes to represent 
both states of each Boolean variable
and thresholds (t-unit nodes) to represent AND and OR relationships present in a node's redescribed look-up table (LUT, Fig. \ref{fig:GRN_representations}d).}

Inference is performed by evaluating the transfer functions that determine a node's state update, which may be written as a logical mapping between possible input vectors to the node's output 
or as Boolean expressions (see Fig. \ref{fig:example_OR}).  Any LUT can be converted to a Boolean expression, and any Boolean expression can be converted to disjunctive normal form (DNF).
In this way, the Boolean expression can be described only by AND, OR, and NOT operators, and the logical satisfaction of any clause guarantees the entire expression to be true.  Klamt \textit{et al.} use DNF of Boolean functions to infer propagation of downstream signals based on the perturbation of certain input nodes \cite{klamt2006methodology}.  Similarly, Wang \textit{et al.} use DNF of Boolean functions to infer cascading failures by removing certain nodes in the network \cite{wang2011elementary}.
Alternatively, the Quine-McCluskey Boolean minimization algorithm \cite{Quine1955truth} can be used to reduce a Boolean expression to its prime implicants \cite{thomas1973boolean}. Marques-Pita \textit{et al.} take advantage of it in 
a process called
schema redescription to remove redundancy from node transfer functions and infer downstream influence of controlled nodes \cite{marques2013canalization}.
This algorithm is also used to reduce transfer functions to DNF 
in the expanded network \cite{zanudo2015cell,yang2018target}.

An additional method for the estimation of 
domains of
influence is the individual-based mean-field approximation (IBMFA) proposed by Parmer \textit{et al.}~\cite{parmer2022influence}. In their work, the
IBMFA is used to
estimate the probability of a node's state given a pinning perturbation of a seed set. 
Parmer \textit{et al.} leverage the IBMFA to optimally identify the minimal sets of nodes able to drive a Boolean network towards fixed-point attractors. However, their method immediately adapts to the estimation of  
domains of
influence.

Despite the availability of these different methods, it is unclear which is the best to determine 
domains of 
influence. 
At the same time, there is a large amount of similarity between the various methods.  All of the graph-theoretic approaches mentioned above, for example, are exact in their description of the network dynamics; the difference lies only in how transfer functions are represented.  However, the specific representations of the transfer functions are important as they determine how much inference can be made in estimating a node's domain of influence.

The goal of the present paper is to fill these gaps of knowledge.
\new{We first introduce a 
framework 
called the generalized threshold network (GTN),
which allows us to cast the graph-theoretic approaches of \cite{wang2011elementary} and \cite{marques2013canalization} in the same formalism, without sacrificing their representation of exact dynamics.}
Next, we show that a simple search 
algorithm
on the GTN can 
be used to 
estimate the domain of influence of a node; this is similar to 
the calculation of the logical domain of influence on the expanded network \cite{yang2018target} and the calculation of pathway modules on the DCM \cite{parmer2023dynamical}.
Finally, we estimate the domain of influence of nodes using the IBMFA~\cite{parmer2022influence}. We test the performance of the various methods on the corpus of biological signaling and regulatory networks obtained from the Cell Collective repository~\cite{helikar2012cell}.
We find that graph representations based on DNF or schema redescription of transfer functions perform very similarly to one another and very similarly to the IBMFA method.  All three methods underestimate the true domain of influence, 
but outperform naive methods based on LUT representations. 
The IBMFA performs somewhat better at recall of node states found within the domain of influence as compared to the other methods, but performs worse in terms of precision.  The computational cost of each method also varies; the IBMFA takes longer than the other methods to run in sparse networks but it runs quicker than the other methods in dense networks.


\section*{Boolean Networks}

A Boolean network $B$ 
is composed of $N$ nodes, each of which has an associated binary state variable $\sigma_i(t) = 0, 1$ at time $t$.  Nodes are connected via directed edges, as defined by an adjacency matrix $A$ with element $A_{ij}$
if node $j$ has a dynamical dependence on 
node $i$
(Fig.~\ref{fig:GRN_representations}a). The network can contain self-loops. We consider synchronous update rules, so that time is represented by a discrete, integer variable. At time $t$, node $i$ updates its state based on the states of its neighbors 
${\mathcal{N}_i}$
at time $t-1$ and the transfer function $F_i$ that uniquely maps every possible combination of the input values to an output state.  This map is called the look-up table, or LUT, of node $i$.  Clearly, the transfer function $F_i$ can also be written as a logical expression; for example, we can use $\sigma_i = \sigma_j \lor \sigma_k$ to specify a logical OR dependency of node $i$ on neighbors $j$ and $k$ (Fig.~\ref{fig:GRN_representations}b).

A network's dynamical configuration at an arbitrary time $t$ is represented by the vector $\vec{\sigma}(t) = [\sigma_1(t), \sigma_2(t), \ldots, \sigma_N(t)]$.  Since we consider synchronous update, where all nodes simultaneously update their state at each time step $t$, the dynamics of the system is deterministic. Also, irrespective of the initial condition $\vec{\sigma}(t=0)$, the network is guaranteed to eventually reach an attractor, either a fixed point or a limit cycle.

In this work, we consider biological signaling and regulatory networks from the Cell Collective repository \cite{helikar2012cell}.
Biological networks are useful case studies to understand node influence in nonlinear systems 
as nodes generally have reversible states and transfer functions are heterogeneous, making analytical approaches difficult.

\section*{Domain of Influence of a Seed Set}

We consider the effect on the system's dynamics of pinning perturbations consisting of imposing and keeping invariant the state of a subset of nodes in the Boolean network. 
We refer to the set of pinned nodes as the seed set, and we indicate it using the notation 
$\mathcal{X} = \{ (i_1, \hat{\sigma}_{i_1}) , (i_2 , \hat{\sigma}_{i_2}), \ldots, (i_{|\mathcal{X}|}, \hat{\sigma}_{i_{|\mathcal{X}|}})\}$,
that is, $\sigma_{i}(t) = \hat{\sigma}_{i} =  0, 1$ for all $(i,\hat{\sigma}_{i})  \in \mathcal{X}$ and for all $t \geq 0$.  Note that, to avoid contradiction, 
$(i, \hat{\sigma}_i ) \in \mathcal{X}$
implies that
$(i, 1 - \hat{\sigma}_i ) \notin \mathcal{X}$.
By contrast, the unperturbed nodes are allowed to change state over time.


If a configuration is sampled at random from the dynamical state space of $B$ at time $t$ given an 
\new{initial set of pinned nodes $\mathcal{X}$,}
each node $i$ has probability 
$P(\sigma_i(t) = 1)$ to be found in the state $\sigma_i = 1$ at time $t$. We refer to this as the activation probability of node $i$ at time $t$~\cite{parmer2022influence}.  
Here, we assume that the state of each node $i$ is initialized with maximally uncertain probability 
$P(\sigma_i(t=0) = 1) = 1/2$
if $i \notin \mathcal{X}$. If $i \in \mathcal{X}$, 
instead $P(\sigma_i(t) = 1) = \hat{\sigma}_{i}$ for $t \geq 0$. This assumption leaves us with a total of  $2^{N-|\mathcal{X}|}$ possible initial configurations, each having the same probability to occur due to the imposed condition on the initial state of the $N-|\mathcal{X}|$ nodes that are outside the seed set. After a transient time period, 
the dynamics started from each 
of these
configurations settles down 
into an attractor. 
\new{As a result, the long-term activation probability of node $i$
converges to a fixed 
value or oscillates, depending on the nature of the attractors being averaged over.}

For networks with $N \leq 10$, we calculate the true activation probabilities by brute-force evaluation over all $R = 2^{N - |\mathcal{X}|}$ possible initial
configurations; otherwise, we sample $R = 100$ randomly chosen initial configurations to obtain an estimate of the ground-truth activation probabilities.
The average state value of a node $i$ at time $t$ based on the $R$ 
sampled initial configurations
is computed as
\begin{equation}
    \overline{\sigma}_i(t) = \frac{1}{R} \, \sum_{r=1}^R \,  \sigma_i^{(r)}(t) \; ,
    \label{eq:av_state}
\end{equation}
where $\sigma_i^{(r)}(t)$ is the state of node $i$ in the $r$-th 
sampled configuration
at time $t$. 

Given the perturbation of the seed set $\mathcal{X}$, the state of another variable $i \notin \mathcal{X}$ in the network may eventually become certain, i.e.,  
$P(\sigma_i(t) = 1) = 0, 1$.
for all $t \geq T$. 
All nodes with deterministic long-term behavior 
compose the domain of influence of the seed set $\mathcal{X}$, i.e., 
\begin{equation}
    \mathcal{D} (\mathcal{X}) = 
    \{ (i, \overline{\sigma}_i(T)) | i \in B \land
    \overline{\sigma}_i(T)=0,1  
    \} \; ,
    \label{eq:doi}
\end{equation}
where $T$ is a finite number of iterations after which the network dynamics is not expected to change, and that can be therefore considered as representative for the long-term dynamics of the network. As in Ref.~\cite{parmer2022influence}, we use $T=10$ to estimate the long-term states of the nodes. \new{This value is in fact sufficient to reach a stationary state under synchronous updating for the set of networks under consideration in this paper~\cite{parmer2022influence}.}
Please note that the set $\mathcal{D}$  
automatically includes  
all elements of
the seed set $\mathcal{X}$.

\section*{Approximating the domain of influence of a seed set}

Determining the ground-truth domain of influence of the seed set $\mathcal{X}$ is generally infeasible, as the task requires to test if nodes reach long-term invariant states for all possible $2^{N - |\mathcal{X}|}$ initial configurations. There are, however, several approaches that can be used to approximate the ground-truth solution in a computationally feasible manner.  Below, we provide a brief description of the various approximate methods considered in this paper.

\subsection*{The individual-based mean-field approximation}

The individual-based mean-field approximation (IBMFA) introduced by Parmer {\it et al.} provides a computationally feasible algorithm to approximate the activation probability of individual nodes in Boolean networks~\cite{parmer2022influence}.
The approximation is inspired by the one generally used in the study of spreading processes on complex networks~\cite{pastor2015epidemic}. The approximation neglects dynamical correlation among state variables to produce predictions in a time that grows as $2^{k_{\max}} \, N$, where $k_{\max}$ indicates the maximum degree of the network. 
The approximation works as follows.
Indicate with $s_i(t)$ the activation probability of node $i$ at time $t$  under the IBMFA. Based on the framing of the problem of identifying the domain of influence of the seed set $\mathcal{X}$, we have $s_i(t=0)=1/2$ if  $i \notin \mathcal{X}$ and $s_i(t\geq 0) = \hat{\sigma}_i$ if $i \in \mathcal{X}$. Then, the activation probability of each node $i \notin \mathcal{X}$ is computed at each time step $t > 0$ according to 
\begin{equation}
    s_i(t) = \sum_{\{n_j: j \in \mathcal{N}_i \}}  
    \delta_{1, F_i ( \vec{n}_{\mathcal{N}_i} ) }
    \, \prod_{j \in \mathcal{N}_i }  
    [s_{j}(t-1)]^{n_{j}} [1 - s_{j}(t-1)]^{1- n_{j}} \; ,
    \label{eq:mf}
\end{equation}
where $\mathcal{N}_i = \{ j^{(i)}_{1}, \ldots, j^{(i)}_{k_i} \} = \{ j \in B 
| A_{ji} = 1 \}$ is the neighborhood of $i$ and $F_i(\vec{\sigma}_{\mathcal{N}_i})$ is the transfer function of $i$ that depends on the network configuration at time $t-1$ restricted to $i$'s neighborhood. We use the IBMFA to estimate the domain of influence of the seed set $\mathcal{X}$ as
\begin{equation}
    \mathcal{D}_{\textrm{IBMFA}} (\mathcal{X}) = 
    \{ (i, s_i(T)) | i \in B \land
    s_i(T)=0,1  \} \; .
    \label{eq:mf_doi}
\end{equation}

\subsection*{Graph-theoretic approximations}

\begin{figure*}
\centering
\includegraphics[width=1.9\columnwidth]{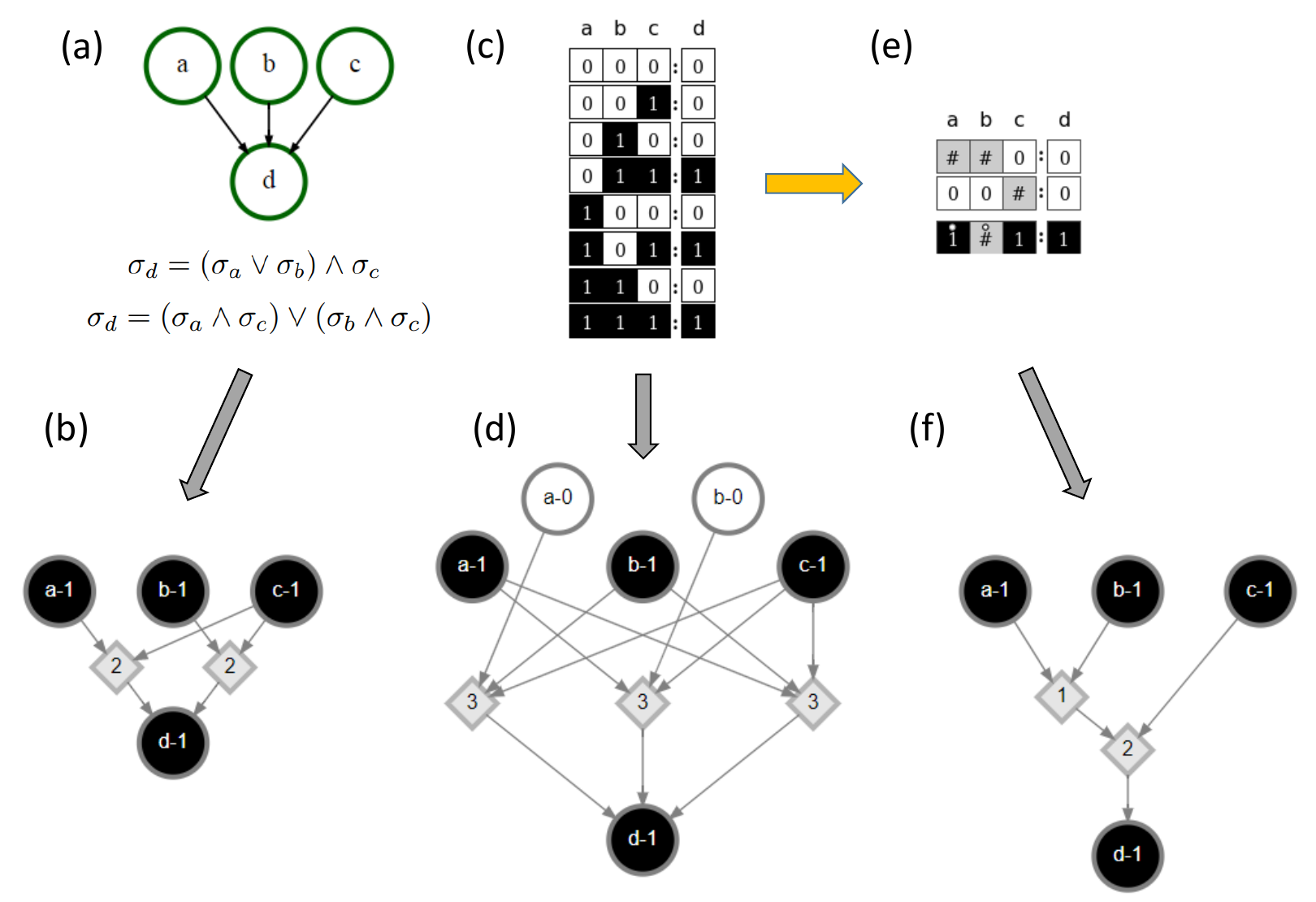} 
\caption{
\textbf{Representations of transition functions and associated generalized threshold networks}. 
(a) A Boolean network composed of four nodes. The network contains three input nodes, $a$, $b$ and $c$. The logical expression
of the transfer function of node $d$ is provided and converted to disjunctive normal form (DNF).
(b) The corresponding generalized threshold network (GTN) of the DNF is displayed as a graph with $6$  nodes and $5$ edges. To keep the visualization compact, we display only the portion of the GTN that concerns the activation of node $d$. (c) The transition function of node $d$ is shown as a look-up table (LUT). (d) The GTN representation of the LUT is displayed as a graph with $9$  nodes and $12$ edges. Also here, we display only the portion of the GTN that concerns the activation of node $d$. (e) The transition function of node $d$ is shown as the two-symbol schema redescription. The wildcard symbol (\#) indicates a node that can have either state value; the position-free symbol ($^o$) denotes that the two indicated state values can switch; that is either $a$ or $b$ may be active to ensure that node $d$ is in the active state, while the state of the other input does not matter. (f) The GTN representation of the two-symbol schema redescription is displayed as a graph with $6$ nodes and $5$ edges. Here too, only the portion of the GTN that concerns the activation of node $d$ is displayed.
}
\label{fig:example_OR}
\end{figure*}

We consider a class of approximations based on static graph representations of Boolean dynamical systems. \new{The various methods are presented within a unified framework based on the so-called 
generalized threshold network (GTN), 
which allows us to study different graph representations of the node transfer functions. 
The GTN is a thresholded network \cite{mcculloch1943logical,marques2013canalization} that represents the entire dynamics 
of a Boolean network $B$. 
As with the 
LIH \cite{klamt2006methodology}, the expanded network \cite{wang2011elementary}, and the DCM \cite{marques2013canalization},
the GTN is also dynamically and logically complete in that every possible dynamical interaction is represented and 
state transitions are unambiguous;
however, it is ambivalent towards the representation of the transfer function used.}

The GTN of a Boolean network $B$ is composed of three sets: the set of  state nodes, the set of  threshold nodes, and the set of  direct edges connecting state and threshold nodes.
If $B$ contains $N$ nodes, its GTN has $2N$ state nodes. Each of the state nodes represents a specific state of the nodes in the Boolean network. The state node with label $i$-$0$ indicates the state $\sigma_i = 0$ of node $i \in B$; the state node with label $i$-$1$ stands for the state $\sigma_i = 1$ of node $i \in B$. 
The set of state nodes is thus functionally equivalent to the set of original and complementary nodes in the expanded network or to the set of s-units in the DCM.
State nodes interact
through threshold nodes; threshold nodes determine the logic that allows for state transitions.
For example, a composite node in the expanded network representing an AND relationship between $k$ inputs is replaced by a threshold node with threshold equal to $k$.
\new{Threshold nodes are used in the same manner as t-units in the DCM.  However, rather than using hyperedges, the GTN uses the alternate representation of Boolean functions in \cite{marques2013canalization} where multiple layers of threshold nodes are possible (as is used, for example, to represent permutation redundancy).  
Thus, the GTN does not restrict how threshold nodes are connected or how many layers of threshold nodes are used, as long as the underlying transition logic is valid,
offering full flexibility in the representation of the transfer function.}

One major limitation of using the LIH or the expanded network is that these network constructions rely on a simplified representation of the transfer functions, such as DNF of the logical expression dictating a node's update.  However, for large Boolean expressions, determining DNF of the expression is infeasible and does not lead to a concise description; additionally, not all networks have their update functions in logical rule format, requiring the logical rules to be determined from the nodes' LUTs. The DCM, by contrast, is constructed via redescribed transfer functions that are in general more concise; however, the process of redescription is also NP-hard and impossible to do for expressions with a large number of inputs. Thus it's unclear how to best represent transfer functions in the optimal way that is both concise and computationally efficient.

The GTN has several advantages. First, it offers a generalization of the expanded network and the DCM where transfer functions are not restricted in their construction. Second, it allows a shared description for different types of dynamical networks, so that expanded networks and DCMs can be compared directly.  Third, it can naturally accommodate variables with any discrete number of values, not just two as in the case of Boolean networks;
further, not all variables need to have the same number of states. Finally, it allows for the exploration of studies on which transfer function representations 
are most useful for predicting dynamics
in different types of networks.

\subsubsection*{Transfer function representations}

The exact structure of a GTN depends on how transfer functions are represented. More precisely, the set of state nodes
is always the same; however, the set of edges and the set of threshold nodes
may change depending on the 
specific choice for the representation of the transfer functions.
There are many such possible representations, and we do not attempt to enumerate them all here. Here, 
we consider a representation based on disjunctive normal form (DNF) as this is used in studies of the LIH and the expanded network \cite{klamt2006methodology,samaga2010computing,zanudo2015cell,yang2018target}.
Also, we  consider a representation based on schema redescription (SR) as this is used in studies of the DCM \cite{marques2013canalization,parmer2023dynamical}. 
Finally, we consider a naive representation that relies only on the LUT of each node without any additional logical reduction.
\new{GTNs based on LUT representations provide worst-case baselines 
in terms of size and performance
compared to GTNs that rely on other more sophisticated and accurate representations.}

To construct a GTN from 
DNF, we first find the DNF 
of the logical update expression of each node 
and also the DNF 
of the negation of the logical update expression. Please note that each network in the Cell Collective repository already has these logical expressions available.
\new{These logical expressions are not reduced from their given form in the Cell Collective; therefore, their DNF may or may not be composed of prime implicants (see SM for details).}
For a given logical expression of node $i$, each disjunctive clause is separated; the state node
for every input in that clause is connected to a threshold node
whose threshold is equal to the number of inputs in the clause. 
The threshold node
is then connected to the state node $i$-$1$ if the logical expression implies $\sigma_i=1$ or to the state node $i$-$0$ otherwise.
For example, in Fig. \ref{fig:example_OR}a, the logical DNF expression dictating the state of node $d$ is
$\sigma_d = (\sigma_a \land \sigma_c) \lor (\sigma_b \land \sigma_c)$. 
There are two separate clauses in this expression, and each one is represented in the GTN using a threshold node.  Each clause has two literals and thus each associated threshold node has threshold equal to $2$ (Fig. \ref{fig:example_OR}b).

To construct a GTN from the SR form, we first find the two-symbol schema redescription of each node's LUT and construct the DCM \cite{marques2013canalization,correia2018cana}. 
All s-units are kept as state nodes,
and all t-units are kept as threshold nodes.
Then, we add other threshold nodes
wherever two edges are fused together to remove all hyperedges in the network.
Thus, the GTN 
transfer functions become equivalent to the intermediate threshold network representation of the canalyzing maps mentioned in \cite{marques2013canalization}.
In Fig. \ref{fig:example_OR}e, the redescribed LUT shows that either $a$ or $b$ can 
be active ($\sigma_{a}=1$ or $\sigma_{b}=1$)
while the state of the other does not matter; in addition, $c$ must also be in state $\sigma_c=1$ in order for $d$ to have state $\sigma_d=1$. The corresponding GTN representation in Fig. \ref{fig:example_OR}f has two threshold nodes: the first indicates that either $a$-$1$ or $b$-$1$ must be present, and the second indicates that the first threshold must be met and $c$-$1$ must be present to reach $d$-$1$.

Finally, to construct a GTN from a LUT form, we first find the LUT mapping of each node's transfer function. 
Then, we split the LUT of node $i$ into rows with output $\sigma_i=1$ and  rows with output $\sigma_i=0$. Next, we create a single Boolean expression for the rows with output $\sigma_i=1$ using OR expressions between each row.
We do the same for all rows with output $\sigma_i=0$. After that, we create AND expressions between each input in each row. Thus, the expression is automatically in DNF 
and can be converted into the GTN representation in the same way as described above for the DNF method. 
In Fig. \ref{fig:example_OR}c, the LUT of node $d$ has three rows that result in state $\sigma_d=1$, and each row has three inputs. This can be converted to the logical expression 
$\sigma_d = (\neg \sigma_{a} \land \sigma_{b} \land \sigma_{c}) \lor (\sigma_{a} \land \neg \sigma_{b} \land \sigma_{c}) \lor (\sigma_{a} \land \sigma_{b} \land \sigma_{c})$, which is automatically in DNF.  As there are three separate clauses, the corresponding GTN representation has three threshold nodes; each clause has three literals and so each threshold node has threshold equal to $3$ (Fig. \ref{fig:example_OR}d).

The LUT representation 
provides upper bounds on the number of threshold nodes 
$M$ and edges $E$
needed in a GTN 
to represent the dynamical system of a Boolean network $B$ with $N$ nodes.
We remind that the number of state nodes in the Boolean network
is $2N$, while the number of threshold nodes $M$
and edges $E$ depends on the number of LUT entries as

\begin{equation}\label{eq:node_upper_bound}
M = \sum_{i=1}^N \, 2^{k_i}
\end{equation}
and
\begin{equation}\label{eq:edge_upper_bound}
E = \sum_{i=1}^N \, (k_i+1) \, 2^{k_i} ,
\end{equation}

where $k_i$ is the degree of node $i \in B$. Eq.~(\ref{eq:edge_upper_bound}) can be derived by noting that there are $2^{k_i}$ rows in the LUT for node $i$; each row requires $k_i$ edges from the neighbors of node $i$ to a 
threshold node, plus an additional edge from the threshold node to
node $i$.
We note an exponential dependence on the nodes' degree for the size of the GTN.  However, the GTN representation can be much more concise by using the DNF or SR forms (see Fig. S1).



\subsubsection*{Identification of the domain of influence}

Given a GTN representation Z of a Boolean network, with Z = SR, DNF or LUT, we estimate the domain of influence $\mathcal{D}_{\textrm{Z}} (\mathcal{X})$ of the arbitrary seed set $\mathcal{X}$  via a breadth-first-search (BFS) algorithm. 
This algorithm is similar to
the one used in Ref.~\cite{yang2018target} to find so-called logical domains of influence. The algorithm works as follows.

We indicate with $r$ the stage of the algorithm, with $\mathcal{Q}_r$ the queue of stage $r$, and with $\mathcal{S}$ 
the set of already visited nodes in the GTN.
We set $r=0$, and we include all elements of the seed set $\mathcal{X}$ in the initial queue, i.e., $\mathcal{Q}_{r=0} = \{ i$-$\hat{\sigma}_i | (i, \hat{\sigma}_i) \in \mathcal{X} \}$. Further, we initialize the set 
$\mathcal{S} = \emptyset$ and the domain of influence $\mathcal{D}_{\textrm{Z}} (\mathcal{X}) = \emptyset$. We then iterate the following instructions:

\begin{enumerate}
    \item  We create an empty queue for the next stage of the algorithm, i.e., $\mathcal{Q}_{r+1} = \emptyset$.
    \item While the queue $\mathcal{Q}_r$ is not empty, we pop one element $e$ out of the queue $\mathcal{Q}_r$. We add $e$ to the set $\mathcal{S}$. If $e$ is a state node, we add the corresponding 
    element
    to the domain of influence, i.e., if $e = i$-${\sigma}_i$ then 
    element $(i, {\sigma}_i)$ 
    is included in $\mathcal{D}_{\textrm{Z}} (\mathcal{X})$. Next, for each neighbor $n$ of $e$ in the GTN, if $n \notin \mathcal{S}$, we consider the following options:
    \begin{itemize}
    \item We add $n$ to $\mathcal{Q}_{r}$ if
        $n$ is a threshold node and its threshold is met by state nodes currently in $\mathcal{S}$.
        
        \item We add $n$ to $\mathcal{Q}_{r+1}$ if
        $n$ is a state node that does not contradict any state nodes already in $\mathcal{S}$.
    \end{itemize}

    \item If $\mathcal{Q}_{r+1} \neq \emptyset$,  we increase $r \to r+1$, and we go back to point 1. Otherwise, we terminate the algorithm.
    
\end{enumerate}

The necessity of having two queues $\mathcal{Q}_{r}$ and $\mathcal{Q}_{r+1}$ is so that state nodes are updated in discrete iterations (i.e., BFS levels).  Threshold nodes, by contrast, are dealt with immediately so that transfer functions can be evaluated before the next stage of the algorithm. 
Note that the two sets $\mathcal{S}$ and $\mathcal{D}_{\textrm{Z}} (\mathcal{X})$ differ since the former is composed of (state and threshold) nodes of the GTN, whereas the latter contains 
nodes of the original graph, along with their corresponding state.

Note that although different GTN representations all give a complete mapping of the network's dynamics, they give different estimates in general for the domain of influence of a seed set $\mathcal{X}$.
For example, in Fig.~\ref{fig:example_OR}, the domain of influence of the seed set $\mathcal{X} = \{(a,\sigma_{a}=1),(c,\sigma_{c}=1)\}$ is 
$\{ a$-1, $c$-1, $d$-$1\}$ for the DNF or SR representations but only $\{ a$-1, $c$-$1 \}$ for the LUT representation.  This is because no logical reduction is done on the node LUTs in this representation, and so inference on the state of $d$ requires knowledge of all three inputs, rather than only two.  As such, performance of this method 
should be a lower-bound for the performance of other representations, such as DNF or SR.
Furthermore, although the DNF and SR representations predict the same domain of influence in this example, note that the SR representation is slightly more concise.

\subsection*{Metrics of performance}


Given a Boolean network and a seed set $\mathcal{X}$, we obtain various approximations of the domain of influence, namely $\mathcal{D}_{\textrm{IBMFA}} (\mathcal{X})$, $\mathcal{D}_{\textrm{DNF}} (\mathcal{X})$, $\mathcal{D}_{\textrm{LUT}} (\mathcal{X})$, and $\mathcal{D}_{\textrm{SR}} (\mathcal{X})$. Also, we find the estimate the ground-truth domain of influence $\mathcal{D} (\mathcal{X})$  via Eqs.~(\ref{eq:av_state}) and ~(\ref{eq:doi}). For compactness of notation, we remove the explicit dependence on $\mathcal{X}$, so that we write $\mathcal{D}_{\textrm{Z}}$ to denote 
the generic approximate set $\mathcal{D}_{\textrm{Z}} (\mathcal{X})$  and $\mathcal{D}$ to denote the ground-truth set $\mathcal{D}(\mathcal{X})$.

We take advantage of multiple metrics to compare the various sets.
Specifically, we compare each approximate solution $\mathcal{D}_{\textrm{Z}}$ against the
ground-truth $\mathcal{D}$ in terms of precision and recall. We determine the set of true positive  defined as $\mathcal{A}_\textrm{TP} = \{  (i, \hat{\sigma}_i) \in \mathcal{D} \land (i, \hat{\sigma}_i) \in \mathcal{D}_\textrm{Z}  \}$,
the set of true negatives as $\mathcal{A}_\textrm{TN} = \{ (i, \hat{\sigma}_i) \notin \mathcal{D} \land (i, \hat{\sigma}_i) \notin \mathcal{D}_\textrm{Z}  \}$, 
the set of false positives as $\mathcal{A}_\textrm{FP} = \{  (i, \hat{\sigma}_i) \notin \mathcal{D} \land (i, \hat{\sigma}_i) \in \mathcal{D}_\textrm{Z}  \}$, 
and the set of false negatives as $\mathcal{A}_\textrm{FN} = \{ (i, \hat{\sigma}_i) \in \mathcal{D} \land (i, \hat{\sigma}_i) \notin \mathcal{D}_\textrm{Z}  \}$.
We compute precision as the ratio $|\mathcal{A}_\textrm{TP}| / (|\mathcal{A}_\textrm{TP}| + |\mathcal{A}_\textrm{FP}|)$, and recall as the ratio $|\mathcal{A}_\textrm{TP}|  / (|\mathcal{A}_\textrm{TP}| + |\mathcal{A}_\textrm{FN}|)$. 
Also, we measure the similarity of the approximation Z with the ground truth using the Jaccard index, i.e., $J_{\mathcal{D}, \mathcal{D}_{\textrm{Z}}} = (\mathcal{D} \cap \mathcal{D}_{\textrm{Z}}) / (\mathcal{D} \cup \mathcal{D}_{\textrm{Z}})$. Such a metric of similarity is used also in the straight comparisons between pairs of approximations, i.e., we rely on $J_{\mathcal{D}_{\textrm{Y}}, \mathcal{D}_{\textrm{Z}}} = (\mathcal{D}_{\textrm{Y}} \cap \mathcal{D}_{\textrm{Z}}) / (\mathcal{D}_{\textrm{Y}} \cup \mathcal{D}_{\textrm{Z}})$ to contrast approximations $Y$ and $Z$.


\section*{Results}

\begin{figure}[!htb]
\includegraphics[width=\columnwidth]{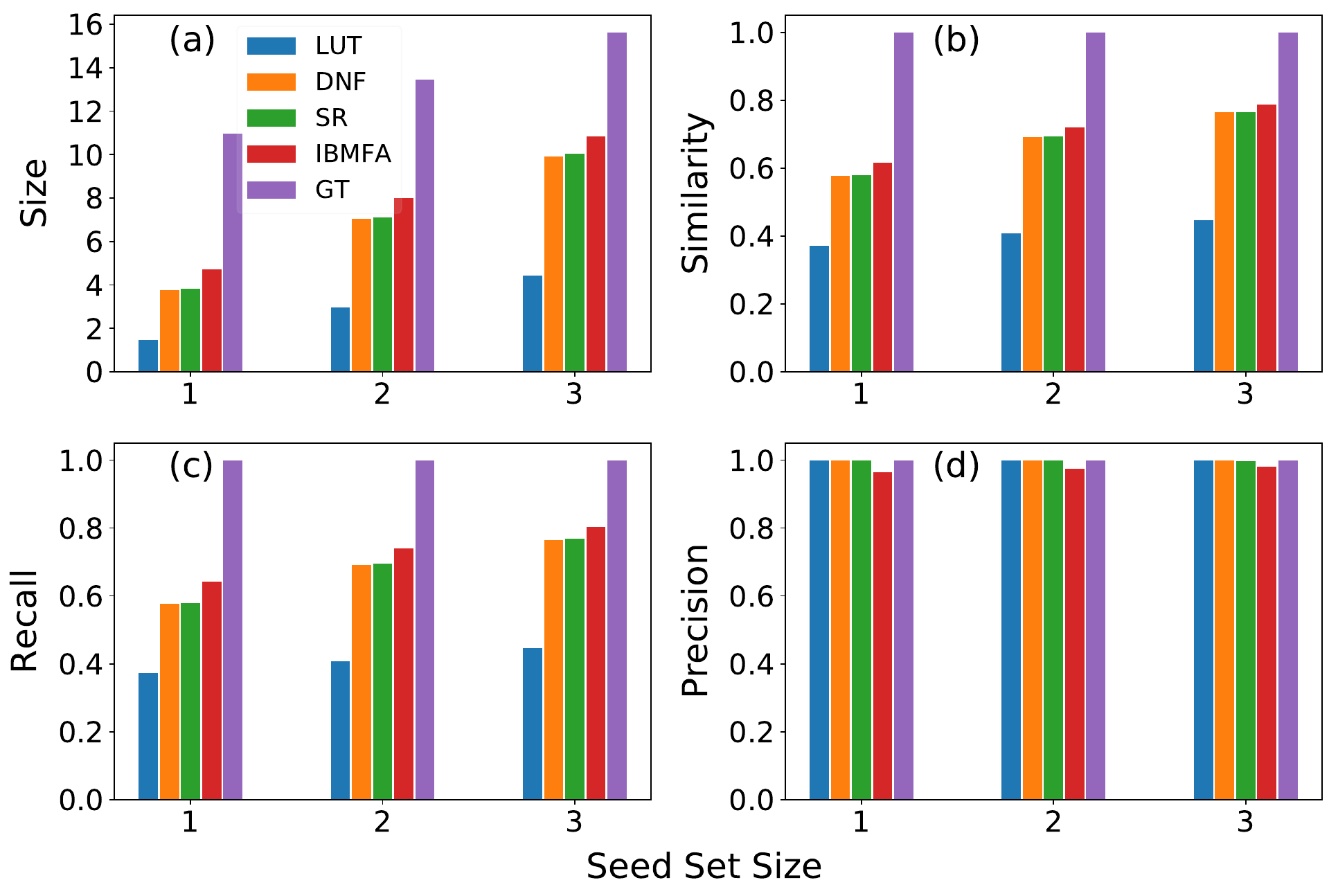} 
\caption{
\textbf{Ground-truth {\it vs.} approximate domains of influence in real-world networks.}
(a) Average size of the domain of influence for networks in the Cell Collective repository. Data are grouped based on the size of the seed set, and displayed results are obtained by taking the average over all sets of a given size included in our analysis. The figure contains results for the ground-truth estimate of the domain of influence (GT), its individual-based mean-field approximation (IBMFA), and its three approximations obtained from the GTN representations based on the transfer functions defined by the node look-up tables (LUT), disjunctive normal form (DNF) of logical functions, and  schema redescription (SR). \new{We remark that GT estimates are obtained by sampling $R=100$ random configurations from the state space of the network. The size of the state space, however, grows exponentially fast with the network size. Thus, the sample used in estimating the GT can be considered sufficiently large only if the network under study is small enough. The sample becomes instead less representative for the state space as the size of the network increases.} 
(b) Average value of the similarity (i.e., Jaccard index) between approximations of the domain of influence and the ground-truth domain of influence.
(c) Same as in panel (b), but for the average recall of approximate domains of influence as compared to the ground-truth.
(d) Same as in panel (c), but for the average precision of approximate domains of influence as compared to the ground-truth.
}
\label{fig:stat_comparison}
\end{figure}

In order to test which methods best elucidate influence on biological signaling and regulatory networks, we create GTN representations of networks from the Cell Collective repository \cite{helikar2012cell}.
For each of these networks, we consider only seed sets of size $1$, $2$, and $3$. Specifically, we consider all possible seeds sets of size $1$ (in a Boolean network with $N$ nodes there are $2N$  possible seed sets of size $1$). We instead randomly sample $1,000$ seed sets of size $2$ and $3$. For each of these sets, we determine the approximate and ground-truth domains of influence. \new{We limit the analysis to seed sets of size smaller than or equal to $3$ for computational reasons,
and because it is unlikely that a researcher will have the ability or desire to control more than a few variables (e.g., proteins or genes) within an actual laboratory setting making control strategies based on larger interventions impractical.}

Comparison between the various approximations of the domain of influence and the ground-truth domain of influence are displayed in Fig.~\ref{fig:stat_comparison}. All methods underestimate the true size of the domain of influence (Fig. \ref{fig:stat_comparison}a). The LUT method performs the worst, while the DNF and SR methods are nearly identical. The IBMFA performs slightly better than the other approximate methods. These results hold for Jaccard similarity and recall as well (Fig. \ref{fig:stat_comparison}b-c): the IBMFA performs the best, while DNF and SR are nearly identical, and LUT is much worse than the others. For precision, by contrast, all GTN-based methods perform very well and only the IBMFA method makes some small mistakes (Fig. \ref{fig:stat_comparison}d).
We also measure the Spearman's rank correlation coefficient, across the entire corpus of Boolean networks in the Cell Collective repository, between the size of the domain of influence $\mathcal{D}_{\textrm{Z}}$ predicted by approximation Z and its ground-truth counterpart $\mathcal{D}$. The above results are again confirmed (Fig.~S2).
The DNF, SR, and IBMFA methods are nearly identical, and they all perform well in ranking the seed sets,
whereas the LUT method performs poorly.

The poor performance of the LUT method is expected, as this method naively creates a representation from the node LUTs without any logical reduction taking place (as happens instead with the DNF or SR methods).
The similarity between the DNF and SR methods is interesting; the two-symbol redescription of the SR method generally reduces transfer functions further than the DNF method is able to.  However, it appears that this does not have a large impact on the results.  Thus, for many networks, it appears that the DNF description is sufficient for good inference of the domain of influence.
Furthermore, it is noteworthy that the IBMFA performs better than the GTN methods on all measures other than precision, even though it approximates node activation probabilities and does not make exact, causal inferences, as the GTN methods do. However, the IBMFA does not have perfect precision, unlike the GTN methods, meaning that sometimes bad inferences are made. 
There is therefore a tradeoff between recall and precision when choosing methods: DNF and SR methods have better precision but worse recall, on average, than the IBMFA.

\begin{figure}[!htb]
\includegraphics[width=.8\columnwidth]{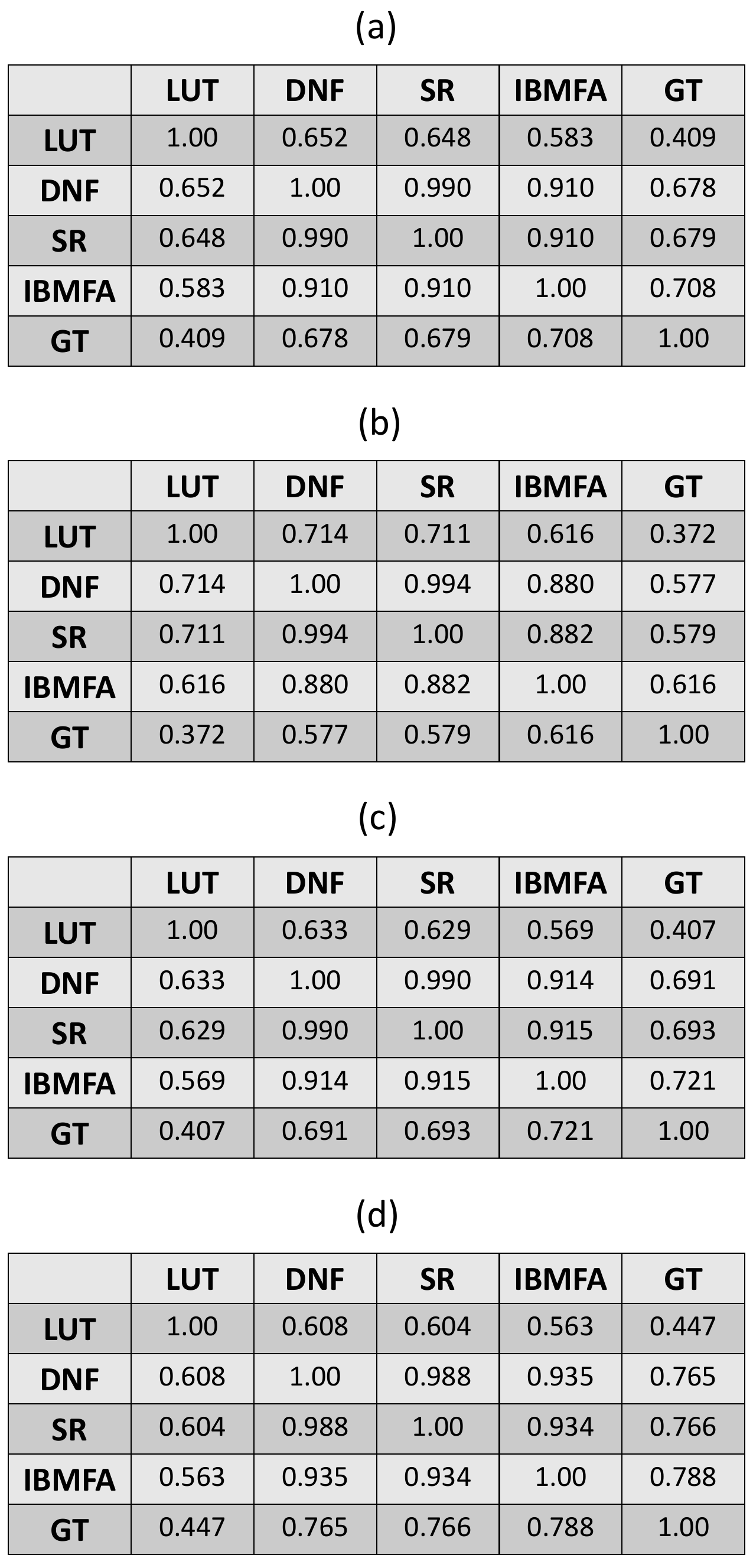} 
\caption{\textbf{Comparison between approximate domains of influence in real-world networks.}
(a) We consider the same set of results as in Fig.~\ref{fig:stat_comparison} and measure the Jaccard index between approximate domains of influence obtained by the various identification methods. Each entry in the tables reports the average value of the similarity score across all networks in the data set, and for all values of the size of the seed set. \new{GT estimates are obtained by sampling $R=100$ random configurations.
}
(b) Same as panel (a), but results are calculated only for seed sets of size $1$.
(c) Same as panel (a), but results are calculated only for seed sets of size $2$.
(d) Same as panel (a), but results are calculated only for seed sets of size $3$.
}
\label{fig:similarity_comparison}
\end{figure}

We measure the similarity between the domains of influence obtained through the various approximations in Fig.~\ref{fig:similarity_comparison}. We see that
SR and DNF generate almost identical predictions; those predictions are also pretty similar to those obtained with the IBMFA method
(see also Fig.~S3). 
By contrast, the LUT method makes predictions of the domain of influence quite different from those of the other methods.

We further analyze the dependence of our results on the network size 
in Fig.~S4. Interestingly, the size of the domain of influence of a seed set increases with the network size, regardless of the size of the seed set (Pearson's correlation coefficient $r=0.46$ for seed set sizes equal to $3$).
This property is valid for the DNF, SR, and IBMFA methods, but not for the LUT method which shows no positive correlation with network size. 
This finding suggests that the LUT method performs worse as 
the network size increases, which is shown also by decreased similarity and recall scores in larger networks. 
However,
the performance of the DNF, SR, and IBMFA methods, as measured by similarity and recall, also decreases as network size increases.

We note an important limitation of our ground-truth estimate of the domain of influence in that the percentage of the dynamical state space sampled by $R=100$ random configurations becomes vanishingly small as $N$ increases (see Fig.~S5).
Under-sampling of the state space can cause our ground-truth predictions to overestimate the size of the domain of influence by predicting false positives, and this could be an alternate explanation as to why the performance of the DNF, SR, and IBMFA methods appears to decrease as network size increases.
It is important, therefore, to verify that our estimated domain of influence accurately reflects the true dynamics of the various networks under study.  Toward this end, we select six networks from the Cell Collective that have size $10 \leq N \leq 14$.  We calculate the true domain of influence for seed sets on these networks via brute-force enumeration of all possible configurations.  We then compare the estimated ground-truth domain of influence $\mathcal{D}_p$ based on sampling a fraction $p$ of the state space to the true domain of influence $\mathcal{D}$. This is done by randomly sampling $R=p 2^N$ initial configurations. We test five values: $p=0.01$, $0.10, 0.25,0 .50$ and $0.75$. 

As the results of Fig.~S6 show, the accuracy of the estimate $\mathcal{D}_p$ depends on the metric being used and the value of $p$.  In this problem, there are no false negatives; therefore, similarity is equal to precision and recall is always 100\%. As such, we show only size and similarity in the figure.
For size and similarity/precision, all $p$ values perform very similarly except for $p=0.01$ which performs worse than the other values.  The value $p=0.10$ also deviates from the true value, although this deviation is small and decreases as seed set size 
is increased.  By contrast, the deviation of $p=0.01$ increases as the seed size increases.  However, even in this case, the similarity/precision is high at about 90\% for seed sets of size $3$.  
Nevertheless, we find that small samples do indeed overestimate the size of the domain of influence and have lower precision (i.e., they have more false positives). This suggests that our results for the ground-truth estimates in Fig. \ref{fig:stat_comparison} may similarly overestimate the size of $\mathcal{D}$ and the number of true positives, 
and this may contribute to the decreased scores of the DNF, SR, and IBMFA methods in terms of size, similarity, and recall.

\begin{figure}[!htb]
\centering
\includegraphics[width=\columnwidth]{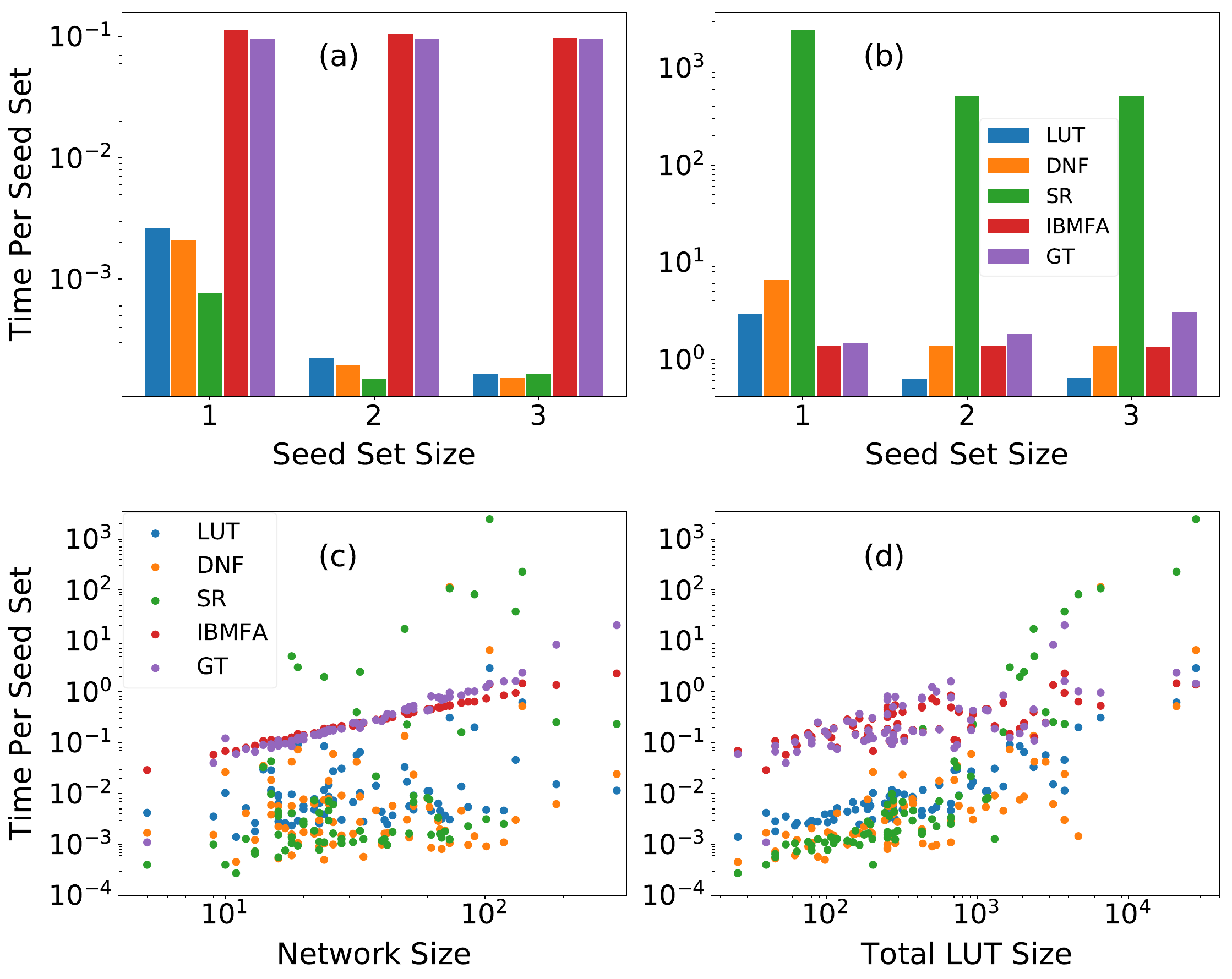} 
\caption{\textbf{Computational time for the estimation of the domain of influence in real networks.}
(a) We measure the time required to estimate domains of influence in the \textit{Drosophila Melanogaster} single-cell segment polarity network \cite{albert2003topology,marques2013canalization}. The size of this network is $N=17$, while its maximum degree is $k_{\max}=4$. Results are averaged over all seed sets of a given size. 
\new{GT estimates are obtained by sampling $R=100$ random configurations.
}
(b) Same as panel (a), but for the Tumor cell migration EGFR \& ErbB signaling network ($N=104$, $k_{\max}=14$) ~\cite{samaga2009logic}.  
(c) Time required to estimate the domains of influence in all networks within the Cell Collective.
Results are averaged over all seed sets of size $1$. Each point in the plot is a network; 
time is plotted as a function of the network size $N$.
(d) Same as in panel (c), but time is plotted as a function of the total size of the look-up tables in the network, i.e., $\sum_{i=1}^N 2^{k_i}$.
}
\label{fig:time_comparison}
\end{figure}

Finally, we compare the computational time required by the various methods to generate approximations of the domain of influence, see Fig.~\ref{fig:time_comparison}. Times are calculated per network over all seed sets of a given size (computations were performed using an Intel Core i5 3.2 GHz processor). 
For the GTN-based approximations, the time necessary to create the GTN is added into the calculation and similarly averaged over the number of seed sets. Depending on the sparsity of the Boolean network, and the representation of the transfer functions, it may take a long time to create a GTN; however, the advantage to using such a graph dynamical approach is that this operation only has to be performed once. Afterwards, the domain set can be approximated in a time that grows linearly with the number of edges of the GTN, 
upper-bounded by
Eq.~(\ref{eq:edge_upper_bound}).
Runtime for the creation of the GTN is especially noticeable for SR graph representations (see Fig.~\ref{fig:time_comparison}).  In networks with low degree, like the \textit{Drosophila Melanogaster} single-cell segment polarity network~\cite{albert2003topology}, the average time 
to approximate domain sets is lower than other methods considered; however, in networks with high degree, like the EGFR \& ErbB signaling network~\cite{samaga2009logic}, the average runtime is much higher for the SR method than for the other methods.

When we consider all networks in the Cell Collective, we see that the time to calculate domains of influence for the IBMFA method grows linearly with the network size (see Fig. \ref{fig:time_comparison}c).  This time is on average greater than the time to find domains of influence using the DNF or LUT methods on a GTN, while the time to approximate domains of influence using the SR method is more variable and is sometimes much greater than for the IBMFA. None of the GTN-based methods are characterized by a clear relationship between computational time and network size. If we consider the total LUT size for each node in the network, however, we see a clear relationship between this quantity and the runtimes of GTN-based approximations, as running time tends to increase exponentially based on total LUT size (Fig.~\ref{fig:time_comparison}d).


\section*{Discussion}

In this paper, we presented results of a systematic analysis 
aiming at comparing the performance of different types of approximate methods in estimating the domain of influence of seed nodes in Boolean networks. The analysis was carried out on a corpus of 74 real-world biological networks from the Cell Collective repository~\cite{helikar2012cell}. Seeds are nodes in the Boolean network with pinned dynamical state. The domain of influence is defined as the set of nodes (seed and non-seed nodes) whose long-term dynamical state become deterministic as a consequence of the external perturbation that pins the state of the seed nodes. 
Approximate methods considered in this paper belong to two classes: (i) graph-theoretic methods, and (ii) mean-field methods. Methods in class (i) are based on representations of the Boolean dynamics into static graphs; methods in class (ii) rely instead on descriptions of average trajectories of the Boolean dynamics where fluctuations are ignored. In spite of the different spirit of the approximation performed, one of the main findings of our systematic study is that methods from the two classes display similar performance, and  
they can perform quite well if the goal is to determine the seed sets that have the greatest influence on a network.  More in detail, all approximate methods underestimate the ground truth, with mean-field approaches having a better recall but a worse precision than the other class of methods. Computationally speaking, graph-theoretic methods are faster than mean-field ones in sparse networks, but are slower in dense networks. 

An important theoretical byproduct of the present study was the introduction of the so-called generalized threshold network (GTN), i.e., a graphical representation of the state space of a discrete dynamical system taking place on a network structure.
The GTN serves as a generalization of the existing approaches by \cite{wang2011elementary} and \cite{marques2013canalization}, but it offers a unified framework that can be applied regardless of the specific representation.
In this paper, we considered three different representations: those based on disjunctive normal form of node logical expressions, those based on schema redescription of node look-up tables, and those based naively on nodes' look-up tables without further logical inference being made.

We stress that the results of this paper are affected by some limitations. \new{First, all our analysis is based on the assumption that the dynamics of a biological network is faithfully represented by the corresponding Boolean model at our disposal. 
If not, the domain of influence will be incorrect as predicted by any method; however, efficiently finding an incorrect domain of influence can help to further improve the model by allowing researchers to spot inconsistencies and modify the underlying dynamical rules to better match the biological reality.  In this sense, the methods here may still be useful for model validation and correction.
Second, our conclusions are valid only for synchronous dynamics, and do not necessarily generalize to other updating schemes, e.g., deterministic asynchronous, stochastic asynchronous, and block deterministic updating schemes. In fact, the fixed points are invariant to the choice of the updating scheme, however the size of the basin of attraction of a fixed point is generally affected by the specific rules of the dynamics at hand~\cite{goles2013deconstruction, aracena2011combinatorics, faure2006dynamical}.
However, previous research suggests that the IBMFA is robust to the order of update in predicting control sets towards particular attractors~\cite{parmer2022influence} and should therefore also be robust in predicting target control.  The graph-theoretic approximations are also robust to the order of update.  As pointed out in Ref.~\cite{yang2018target}, the predicted domain of influence will have prefect precision as long as the updating scheme preserves the level order of the breadth-first search on the GTN.}

\new{Even under the assumptions that only one high-fidelity Boolean model exists for a given biological network and that dynamics is synchronous, additional limitations are present.}
For example, our estimates of the ground-truth domain of influence of a seed set depend on sampled configurations that constitute only a small fraction of the actual state space, therefore leading to the appearance of false positives in the ground-truth estimates of the domain of influence. Although our results were confirmed also in small networks where ground-truth estimates are exact (see Fig.~S6c-d),
some of the gaps in size, similarity, and recall seen by the methods tested here may be due to this systematic bias.  
Another major limitation is that these methods were tested only on biological networks of moderate size from the Cell Collective repository, \new{and we considered seed sets of maximum size equal to three}. It is unclear how different methods would perform in larger and/or non-biological networks, \new{and for larger seed sets}.  Further research is needed for this purpose.
Finally, we note that we relied on pre-existing logical expressions in the Cell Collective for the DNF method, and that such expressions are not in general available for all networks.  However, as an alternative, it is possible to find disjunctive normal form of the prime implicants of node LUTs using the Quine-McCluskey algorithm \cite{Quine1955truth}; we find that both methods for finding DNF of node expressions make nearly identical predictions (see Fig.~S7), suggesting that the results seen here are also valid for networks that do not have such pre-existing logical expressions available.

Despite such limitations, this work is a step towards understanding how the behavior of Boolean networks can be effectively and efficiently predicted,
\new{which is essential in target control problems.}
This work further elucidates the type of strategy to be used depending on the network, i.e., sparse {\it vs.} dense, and/or the specific application at hand, i.e., when recall is favored over precision or {\it vice versa}.

\acknowledgements
{
  The authors thank L.M. Rocha for comments and suggestions on the paper.
This project was partially supported by the Army Research Office under contract number W911NF-21-1-0194 and by the Air Force Office of Scientific Research under award number FA9550-21-1-0446. The
funders had no role in study design, data collection and
analysis, the decision to publish, or any opinions, findings,
and conclusions or recommendations expressed in the
manuscript.}

\new{
\section*{Data availability}
Network data can be retrieved at \url{https://cellcollective.org} and
\url{https://github.com/rionbr/CANA/tree/master/cana/datasets/cell_collective}.  The 74 networks we used for our analysis are listed 
at \url{https://github.com/tjparmer/node_influence/blob/main/list_of_networks_analyzed.txt}.

\section*{Code availability}
The code developed for this paper is made available at \url{https://github.com/tjparmer/node_influence}.
}

\bibliographystyle{unsrt}



\end{document}